\documentclass[times, twoside]{zHenriquesLab-StyleBioRxiv}
\usepackage{blindtext}
\usepackage{subfig}
\usepackage{amssymb}
\usepackage{dsfont}
\usepackage{multirow}
\leadauthor{Yiming Fan}

\begin{document}

\title{Attention-aware contrastive learning for  predicting T cell receptor-antigen binding specificity}\shorttitle{Contrastive learning for predicting TCR-antigen binding}

\author[1,\dag]{Yiming Fang}
\author[1,\dag]{Xuejun Liu}
\author[1,$\ast$]{Hui Liu}

\affil[1]{School of Computer Science and Technology, Nanjing Tech University, 211816, Nanjing, China}

\maketitle

\begin{abstract}
It has been verified that only a small fraction of the neoantigens presented by MHC class I molecules on the cell surface can elicit T cells. The limitation can be attributed to the binding specificity of T cell receptor (TCR) to peptide-MHC complex (pMHC). Computational prediction of T cell binding to neoantigens is an challenging and unresolved task. 
In this paper, we propose an attentive-mask contrastive learning model, ATMTCR, for inferring TCR-antigen binding specificity. For each input TCR sequence, we used Transformer encoder to transform it to latent representation, and then masked a proportion of residues guided by attention weights to generate its contrastive view. Pretraining on large-scale TCR CDR3 sequences, we verified that contrastive learning significantly improved the prediction performance of TCR binding to peptide-MHC complex (pMHC). Beyond the detection of important amino acids and their locations in the TCR sequence, our model can also extracted high-order semantic information underlying the TCR-antigen binding specificity. Comparison experiments were conducted on two independent datasets, our method achieved better performance than other existing algorithms. Moreover, we effectively identified important amino acids and their positional preferences through attention weights, which indicated the interpretability of our proposed model.
\end{abstract}

\begin{keywords}
Contrastive learning | T cell receptor | Attention mechanism | Neoantigen | TCR-antigen binding | Transformer
\end{keywords}

\begin{corrauthor}
hliu\at njtech.edu.cn
\end{corrauthor}

\section*{Introduction}
T lymphocytes play a vital role in adaptive immunity~\citep{rf1,rf2}. The recognition and binding to specific antigens presented on the surface of cells is prerequisite to elicit cytotoxic T cells ~\citep{rf3,rf4}. T cells rely on the T cell receptors (TCRs) to recognize antigenic peptides presented by the major histocompatibility complex (MHC) located on the surface of antigen-presenting cells. Human TCR repertoire diversity reach $10^{16}$ to $10^{18}$ so that T cells can potentially recognize a huge number of antigens. Previous studies have shown that TCR-antigen binding specificity~\citep{rf5,rf6,rf7} is mainly determined by the heterodimer composed of two $\alpha$ and $\beta$ peptide chains~\citep{rf21}. This variability in specificity stems mostly from plasticity of three complementarity-determining region (CDR) loops (CDR1-3) of both TCR $\alpha$- and $\beta$-chains, among which CDR3 is responsible for binding to antigenic peptides~\citep{rf1,rf2}. So, CDR3 sequence is highly diverse~\citep{rf8} and the major determinant of antigen binding specificity.  Accurate and quantitative estimation of CDR3 diversity and clonal expansion directly reflect the immune response status of T cells to specific antigen stimulation~\citep{rf9}. TCR-antigen binding specificity is a key factor for the evaluation of therapeutic effect in immunotherapy, such as PD-L1/PD-1 inhibitors, adoptive T-cell immunotherapy and tumor vaccine design. However, the enormous diversity of TCR repertoire makes it a forbidding difficulty for wet-lab experiments to screen out TCR-epitopes binding specificity~\citep{rf10}. There is a pressing need for development of prediction method to predict TCR binding specificity of antigen, which would greatly complement wet-lab experiments.

Several  databases with large collection of experimentally validated TCR-peptide interactions, such as VDJdb~\citep{rf12}, IEDB~\citep{rf13}, and McPAS~\citep{rf14}, greatly facilitated the development of computational prediction models. A variety of deep learning-based methods have been developed to predict TCR-antigen binding specificity and acquired encouraging performance~\citep{rf11,rf15}. For example, Springer et al.~\citep{rf10} developed an ERGO model based on natural language processing (NLP) model, using a dictionary of TCR-peptide interactions to predict the binding of TCR and antigenic peptide; Jurtz et al.~\citep{rf16} developed the NetTCR model based on convolutional neural networks to identify TCRs potentially bound by homologous peptides, making advantage of a large number of non-binding TCR sequences. Lu et al.~\citep{rf17} established a transfer learning-based model pMTnet to predict binding of TCR to neoantigen presented by MHC I complex. Jun et al. ~\citep{rf19} proposed a multi-instance learning MIL framework, BERTMHC, to predict antigen binding and presentation by MHC II proteins; Gielis et al. ~\citep{rf34}developed the web tool TCRex to predict interactions between TCR and epitopes stemmed from multiple types of cancer or viral. These methods have demonstrated the effectiveness of computational prediction of TCR-antigen binding specificity.

Recently, self-supervised representation learning has achieved substantial progress in natural language processing~\citep{rf22}. The milestone BERT model~\citep{rf24}, a pre-trained deep bidirectional language representation model based on Transformer \citep{rf23,rf24}, achieved excellent performance on several downstream tasks. Self-supervised contrastive learning~\citep{rf25} has caught much attention because of its better generalizability in multiple fields. Contrastive learning generated two different views~\citep{rf37} from the original sample, and trained a deep representation network to minimize the contrastive loss, which encouraged to output similar representations for the augmentations (views) of the same image but dissimilar representations for those generated from other images. Quite a few contrastive learning methods have been proposed, such as SimCLR~\citep{rf25}, MoCo~\citep{rf26}, BYOL~\citep{rf31}, SimSiam~\citep{ref36}. They run pre-training on large-scale unlabeled dataset, transferred to specific downstream tasks and achieved superior performance and robustness~\citep{rf26,rf27,rf30}. For example, Wang et al.~\citep{PSSM} overcame the drawback of low-quality Position-Specific Scoring Matrix (PSSM) matrix due to poor sequence homology, by running contrastive learning-based feature extraction. Their model achieved great performance improvements on multiple extremely low-quality cases.

The success of contrastive learning often relies on the quality of contrastive views.  In computer vision, the most commonly used method to generate contrastive views is image augmentation, such as image flipping, random cropping and random Gaussian blur. In graph neural networks, new views were often generated by removal of some nodes, edges or subgraphs. However, all these methods for contrastive view generation pose random perturbation to the original samples. They do not consider the importance of different components of the samples, which may lead to the failure of capturing important information and poor robustness. The feature extraction process spent on non-critical components also results in low computational efficiency.

In this paper, we propose an attention-aware contrastive learning model, ATMTCR, for predicting the binding specificity between TCR and pMHC complex. For each TCR sequence, a proportion of residues were masked under the guidance of attention weights, thereby the contrastive view was generated. In particular, we explored and compared different masking strategies, including masking residues with max- or min- attention weights or roulette probability. We performed exhaustive experiments and demonstrated that contrastive learning significantly improves the predictive performance of TCR-antigen binding. Performance comparison experiments were conducted on two independent TCR-antigen binding datasets, and ATMTCR achieved better performance than other existing algorithms. Moreover, we observed the attention weights help to reveal important amino acids and their positional preferences through, which indicated the interpretability of our proposed model.

To the best of our knowledge, we were the first to apply contrastive learning for pre-training on large-scale TCR sequences. The obtained TCR sequence representations greatly improved downstream multiple tasks. In addition, our proposed attention masking generates different views by minimizing the contrastive loss of the two views, so that the model focuses on key positions and specific amino acids in the sequence, and successfully extracts high-order semantic information from the sequence.

\section*{Materials and Methods}\label{sec2}

\subsection*{Data resource}\label{subsec1}
\begin{figure}[t]
        \centering\includegraphics[scale=0.65]{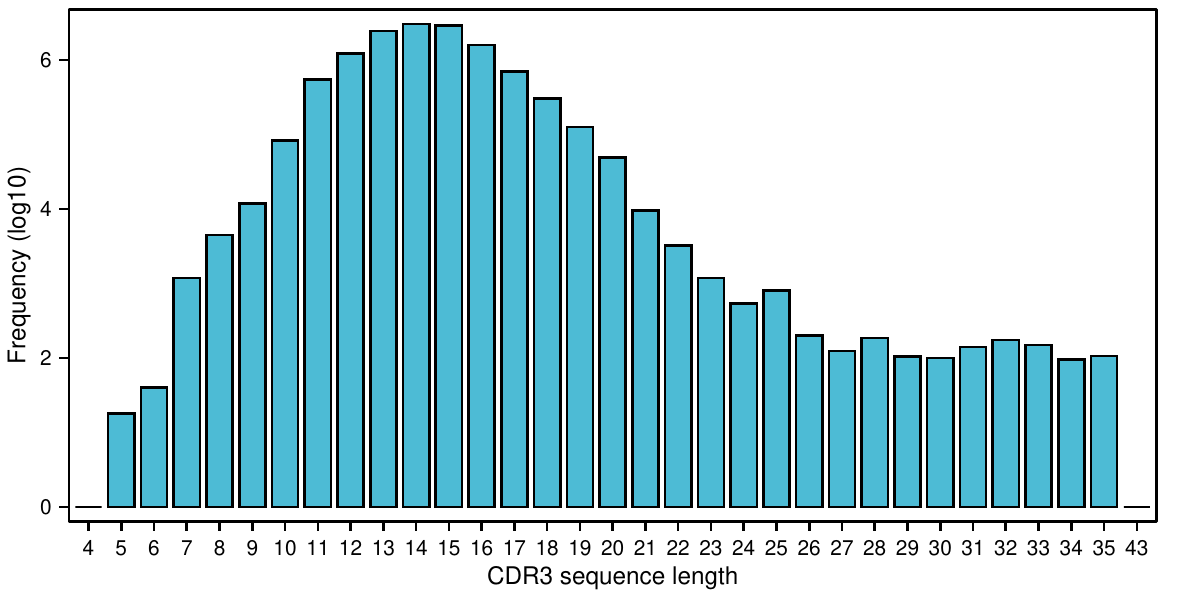}
        \caption{Frequency distribution of TCR CDR3 sequences with respect to length}
        \label{fig:cdr3_len}
\end{figure}

For TCR sequence representation learning, we downloaded 13,148,345 CDR3 sequences from TCRdb~\citep{rf20} database, which collected CDR3 sequences from different cancer types and tissues. Figure \ref{fig:cdr3_len} showed the frequency distribution with respect to CDR3 sequence length. The length ranged from 5 to 35 amino acids. Most CDR3 sequences has 14$\pm$5 amino acids, but a long-tail occurrence can be observed. Although these sequences have no labels of binding to antigens, the large scale of CDR3 sequences increased the diversity of samples, thereby helped to built a CDR3 sequence encoder with strong generalization ability.

For the downstream task of predicting TCR-antigen binding specificity, we used the datasets released by Lu~\citep{rf17} et al. There were two independent datasets. The first includes 32,044 TCR-pMHC bindings between 28,624 unique CDR3 sequences and 428 antigenic peptide presented by 64 MHC alleles. The second includes 619 pMHC/TCR bindings between on 272 unique CDR3 sequences and 224 antigenic peptides presented by 24 alleles. For simplicity, two datasets were referred to as large and small set, respectively.

\subsection*{Model framework}\label{subsec1}
\begin{figure*}[t]
        \centering\includegraphics[scale=0.75]{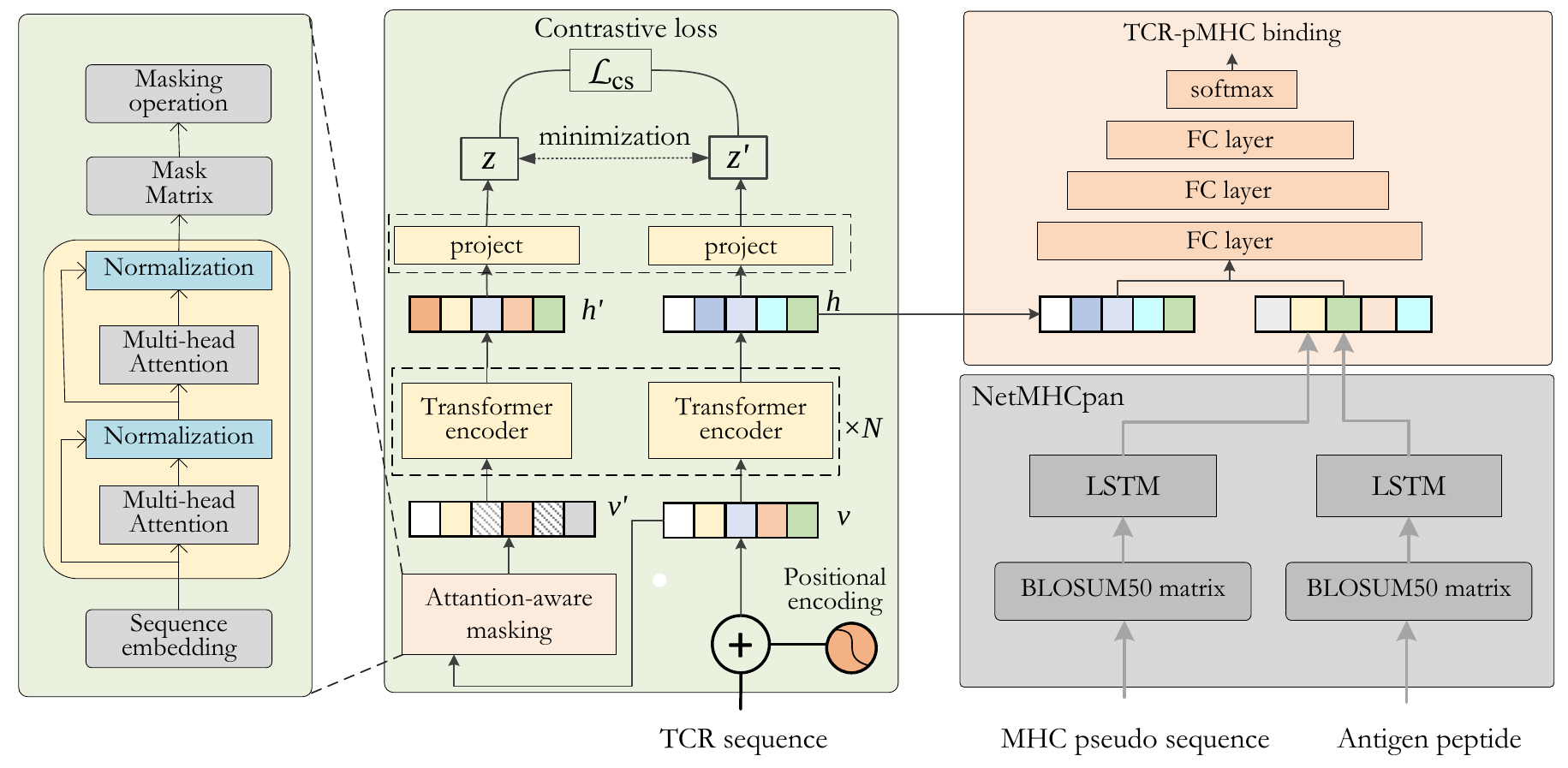}
        \caption{Illustrative flowchart of ATMTCR learning framework, including two stages: contrastive learning-based pretraining (light green box) and transfer learning to TCR-pMHC binding prediction (light red box).  The gray box represented the process for antigen-MHC complex representation learning using NetMHCpan\citep{rf33}. The dotted-line boxes on the transformer and projector represented shared parameters between them, respectively. }
        \label{fig:flowchart}
\end{figure*}

We designed a contrastive learning-based model for the prediction of  TCR-antigen binding
specificity. The flowchart in Figure \ref{fig:flowchart} illustrated that our learning framework included two stage: pretraining and transfer learning. Rather than random masking commonly used to
generate contrastive views, we developed an attention-aware masking module to generate
positive counterpart views. By minimization of the contrastive loss, the network is self-
trained to continuously improve its discrimination ability to distinguish positive pairs from
negative samples. For downstream tasks, we used NetMHCpan~\citep{rf33} to obtain the
representations of  peptide-MHC complexes. By concatenation with the CDR3 sequence
representations derived from contrastive learning, we fed them into two fully-connected
layers to predict the binding of  TCR and pMHC complex.

\subsection*{Transformer-based TCR sequence representation}\label{subsec1}
First, word embedding was used to transform each CDR3 sequence to latent representation by random initialization following normal distribution N(0,1). Formally, the input CDR3 sequence was encoded as a sequence of tokens: $\mathbf{x}=\{x_1,..., x_L\}$, and each token $x_i$ was embed into d-dimension latent space $R^d$. So, we have $\mathbf{x}\in R^{L\times d}$. To make advantage of the position information of each amino acid, we used the sine and cosine encoding to obtain its positional embedding. For the sequences with length less than predefined length $L$, they were aligned by padding a few special character. Finally, the word embedding and the positional embedding were summed and fed to the transformer block.

The transformer block applied multi-head attention mechanism to model the interaction of each amino acid pairs in the input sequence.  For each pair of amino acids $i$ and $j$, the encoder learned an attention score $w_{ij}$>0 with constraint$ {\sum_{j} w_{ij}=1}$. The attention scores are computed from the normalized dot product of query vectors and key vectors followed by a softmax operation. The output of a self-attention layer is a weighted sum of the value vector by the attention weights. The operations of a self-attention layer written in matrix form are as follows:
\begin{equation}\label{equ:attention}
  W=softmax\left(\frac{QK^T}{\sqrt{d_k}}\right)V
\end{equation}
where $d_k$ is the dimension of the key vectors, $Q$ is a query matrix, $K$ is a key matrix and $V$ is a value matrix. The query, key and value matrices in this case are different trainable linear projections of the layer input. Taking unlabeled protein sequence as input, the transformer encoder was trained to learn expressive and discriminative latent representations. The transformer was followed by a three-layer multilayer perceptron (MLP) and activation function ReLU as the encoder to obtain hidden features.

\subsection*{Attention-aware masking for augmentation}\label{subsec1}
An attention-aware masking module was designed to generate contrasting views, whose architecture resemble the transformer encoder.  For each pair of amino acids $i$ and $j$, the encoder learned an attention score $w_{ij}>0$. The attention scores are computed from the normalized dot product of query vectors and key vectors followed by a softmax operation. The output of a self-attention layer is a weighted sum of the value vector by the attention weights. In our study, two attention layers were used, and each layer contained two-head self-attentions. The outputs from each self-attention heads are concatenated to give a final continuous vector for each amino acid.

\begin{figure}[t]
        \centering\includegraphics[scale=0.7]{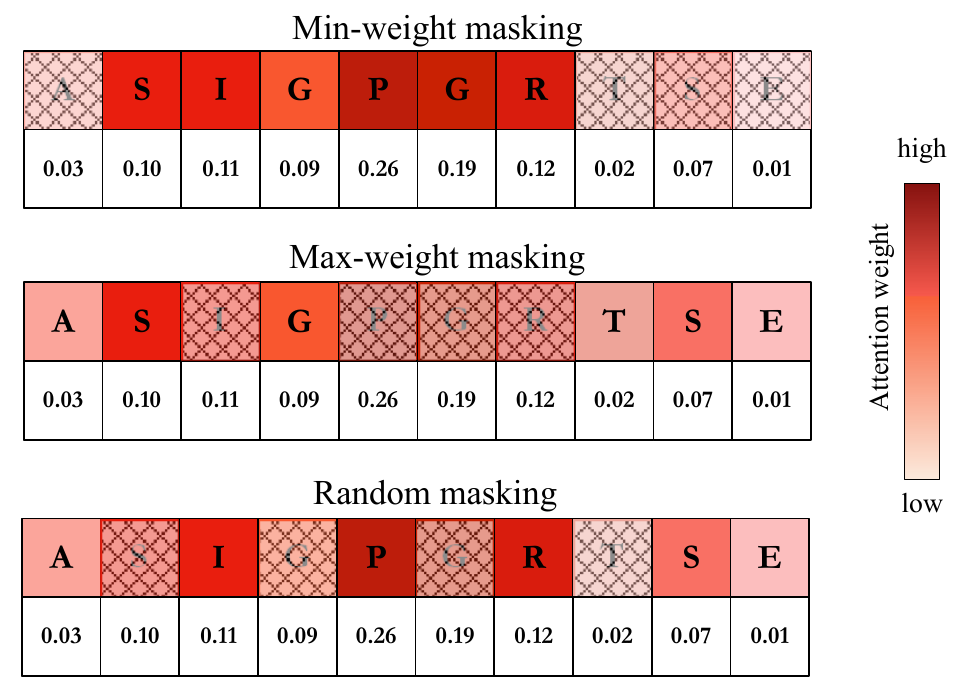}
        \caption{Illustrative diagram of three masking strategies for the generation of contrastive view}
        \label{fig:masking}
\end{figure}

For each input CDR3 sequence, we got an attention matrix $W$, and then transformed it to a vector by computing $w_i=\sum_{j=1}^{d}w_{ij}$ so that the attention weight of amino acid $i$ was obtained. As shown in Figure \ref{fig:masking}, given the attention vector, we generated augmented view for input CDR3 sequences by masking a proportion of residues. To explore the effect of masking different residues, we tried three masking strategies:
\begin{enumerate}
\item Max-attention masking: $r\%$ residues with the largest attention weights were masked. This produced an augmented view with greatest difference to the current view of the input CDR3 sequence each time.

\item Min-attention masking: $r\%$ residues with the smallest attention weights were masked. This produced an augmented graph with least difference to the current view of the input CDR3 sequence each time.

\item Random masking: $r\%$ residues was randomly selected and masked. This masking strategy was commonly used in previous studies, we included it for comparison.
\end{enumerate}
The attention-aware masking to produce augmented view help to reveal important amino acids. Also, it enriched the diversity of negative samples that are helpful to learn expressive and robust representations.

\subsection*{Contrastive learning}\label{subsec1}

The goal of contrastive learning is to capture the data structure from large-scale unlabeled data. In the latent space, contrastive learning learn to maximize the similarity of the representations of positive contrastive views, while minimize their similarity to negative samples. Therefore, the pretrained transformer encoder output discriminative representations~\citep{rf32} for different instances. Formally, the input CDR3 sequences and its augmented view were embed into vectors $h_i$ and $h'_i$, which were then mapped to $z_i$ and $z'_i$ by a nonlinear projector. Next, the similarity $sim(z_i, z'_i)$ between two projected view $z_i$ and $z'_i$ was computed. We adopted the we normalized temperature-scaled cross entropy (NT-Xent) as the contrastive loss function:
\begin{equation}\label{equ:loss}
  \mathcal{L}_{cs}=\log\frac{\exp(sim(z_i,z'_i)/\tau)}{\sum_{k=1}^{2N}{\mathds{1}_{[k\neq i]}\exp(sim(z_i,z_k)/\tau)}}
\end{equation}
where $\mathds{1}_{[k\neq i]}\in\{0,1\}$ is an indicator function evaluating to 1 iff $k=i$, $\tau$ denotes a temperature parameter; $N$ is the number samples in a mini-batch. In our study, the cosine distance was used to evaluate the similarity of two views of same instance.

Note that a pair of contrastive views were transformed by the same transformer encoder, as well as the projector. The attention-aware masking to produce augmented view help to reveal important  amino acids. Also, it enriched the diversity of negative samples that are helpful to learn expressive and robust representations.

\subsection*{Transfer to TCR-pMHC binding prediction}\label{subsec1}
Once the representations of TCR CDR3 sequences were obtained via pre-training stage, we transferred them to the downstream tasks. Since this study focused on the TCR-pMHC binding specificity, we adopted the popular NetMHCpan [33] algorithm to encode the MHC I and antigenic peptide complex. Thereafter, the TCR and pMHC embeddings were concatenated and fed into four fully-connected layers. The first three layers used ReLU activation function, and finally use the sigmoid activation function for classification.

During transfer learning stage, we froze the weights of the transformer encoder and tuned only the parameters of the fully-connected layers. The cross-entropy loss was applied for all classification task, and the learning rate is set to 1e-4. Adam optimizer was used and batch-size was set to 450. The performance of TCR-pMHC binding prediction was evaluated by ROC-AUC. The early stopping and dropout strategies were used to prevent overfitting. Each dataset used for TCR-pMHC binding prediction was split into training, validation and test datasets in a ratio of 8:1:1. The pretrained model was fine-tuned on the training set and validated on the validation set.

\section*{Results }
\subsection*{Contrastive learning improved generalizability}
\begin{figure}[htbp]  
	\captionsetup{labelformat=simple, position=top}
	\centering
	\subfloat[Performance comparison on \textit{small} dataset]
	{
		\begin{minipage}[b]{\columnwidth}
			\centering
			\includegraphics[width=4cm]{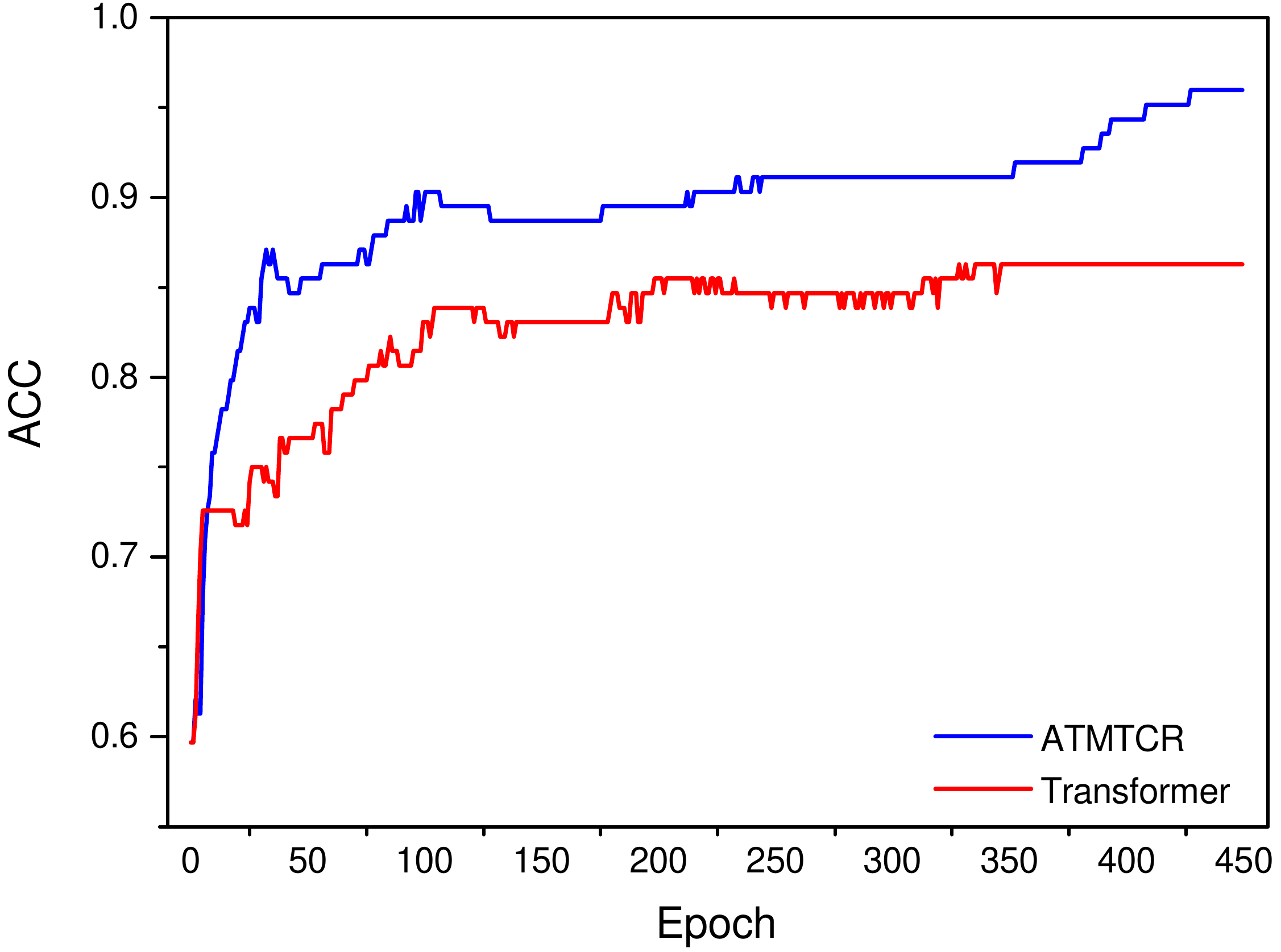}
            \includegraphics[width=4cm]{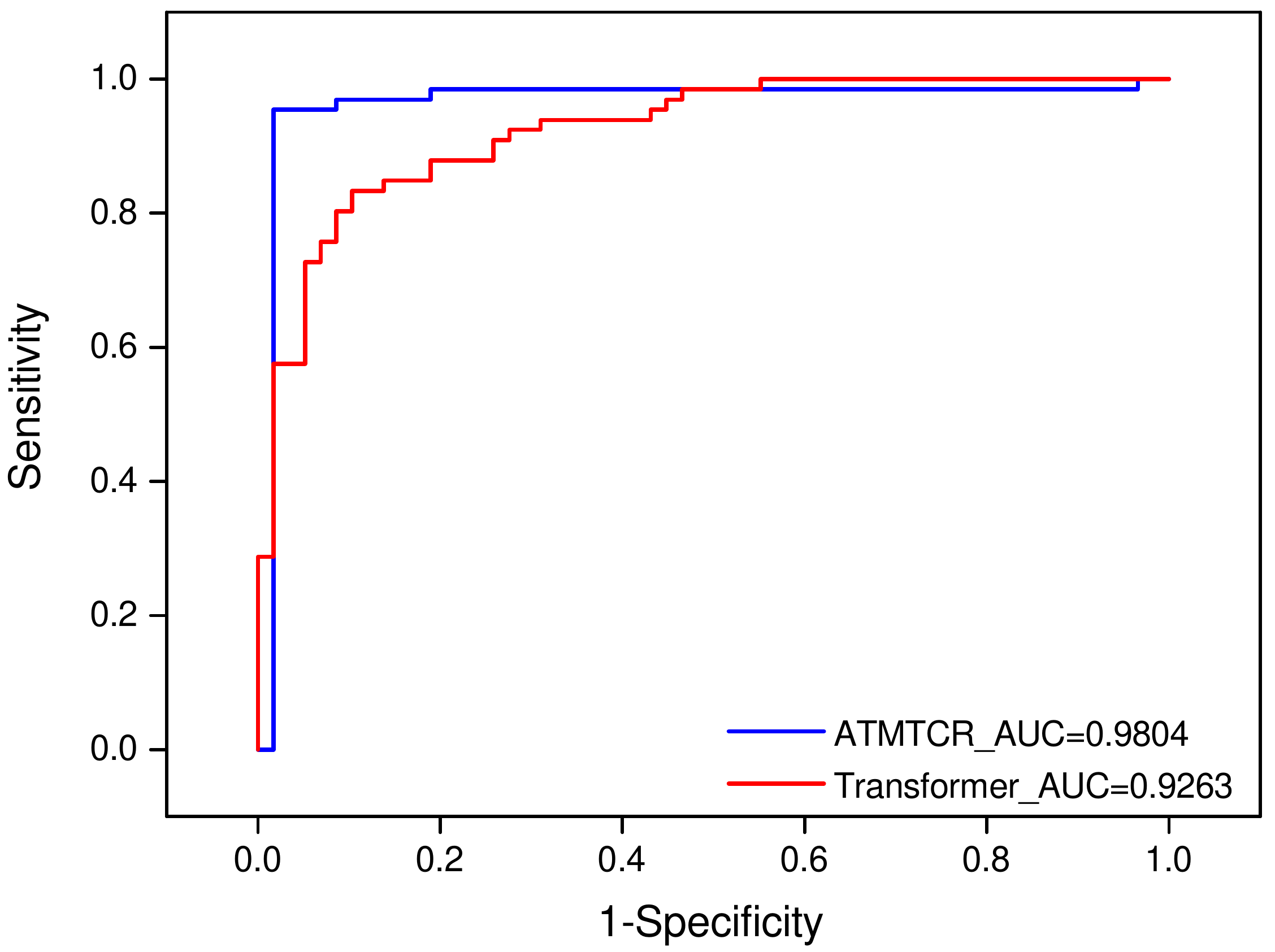}
	\end{minipage}}\\

	\subfloat[Performance comparison on \textit{large} dataset]
	{
		\begin{minipage}[b]{\columnwidth}
			\centering
			\includegraphics[width=4cm]{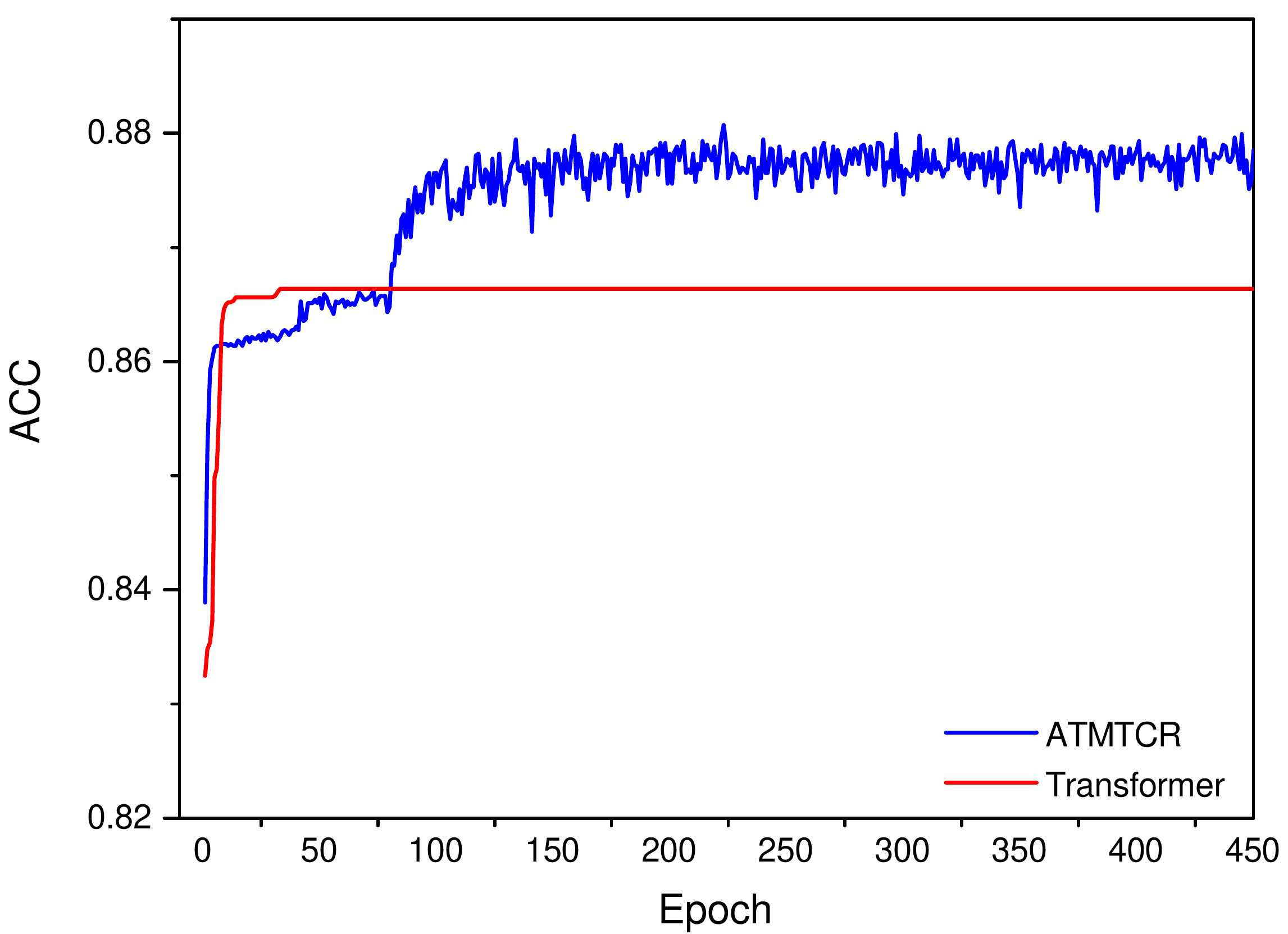}
            \includegraphics[width=4cm]{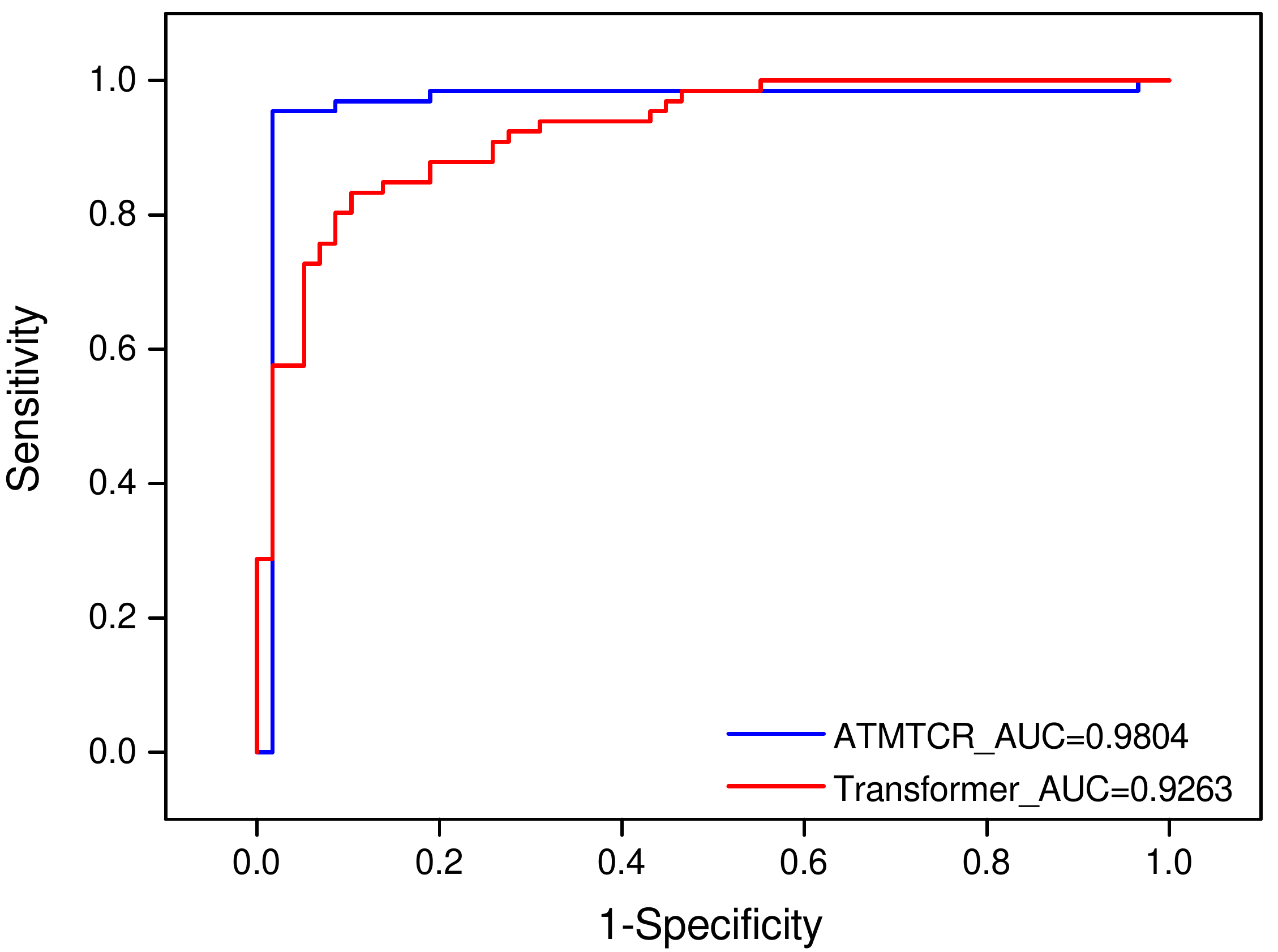}
	\end{minipage}}
	\caption{Performance comparison of contrastive learning-based pretraining to fully-supervised learning evaluated on two datasets}	\label{fig:auc}
\end{figure}

To verify contrastive learning-based pre-training improved the performance of downstream prediction tasks, we compared the proposed model with a fully-supervised learning model without pre-training. For this purpose, a transformer block was used to encode the CDR3 sequences, and directly feed to fully-connected layers. This deep network was trained using only the TCR-pMHC binding dataset. The NetMHCpan released by Nielsen \citep{rf30}, $M$ was used to encode pMHC complexes. The comparison experiments were carried out on two datasets independently, and the results are shown in Figure \ref{fig:auc}. ATMTCR performed much better than the model without pre-training on both datasets. In particular, on the small dataset, ATMTCR achieves 0.95 accuracy, which was $10\%$ higher than the fully-supervised model. This showed that our contrastive learning-based model obtained better generalization ability, thereby achieved excellent performance when transferring to downstream task by fine-tuning on small-size data sets. The performance of supervised learning models was limited due to the insufficiency of training data. From this, we can draw the conclusion that our contrastive learning effectively learns CDR3 sequence features and improves the prediction accuracy of TCR-pMHC binding when applied to downstream tasks.

\subsection*{Attention-aware masking performed better than random masking}\label{subsec2}
Next, we wanted to clarify which was the best masking strategies for generation of contrastive views. So, we tested the effect of different masking strategies on TCR sequence representation learning and downstream tasks. Note that pre-training is all performed on the same unlabeled  dataset. We run the min-weight, max-weight, and random masking in pretraining stage independently, and then transferred to downstream tasks. Table \ref{table:1} showed the ROC-AUC and accuracy values on two downstream datasets.

We found that the min-weight masking outperformed the other two methods, by achieving 0.98 and 0.90 ROC-AUC values on two datasets, respectively. However, random masking and max-weight masking performed inconsistently on the two datasets. The results implied that masking the amino acids with low attention weights was more conducive for the model to concentrate on important amino acids, thereby generating more informative latent representations. Also, we observed that, regardless of the masking strategy, the pretrained model always achieves better performance compared to the fully-supervised model without pretraining (also see Figure \ref{fig:auc}).
\begin{table}[]
    \centering
    \caption{Performance comparison among three masking strategies on two datasets}
    \label{table:1}
    \resizebox{\linewidth}{!}{
    \begin{tabular}{c|c|c|c|c|c}
    \hline
        \multicolumn{2}{c|}{ } & \multicolumn{2}{c|}{Small dataset$(\#619)$} & \multicolumn{2}{c}{Large dataset$(\#32044)$}\\
        \hline
        \multicolumn{2}{c|}{Masking strategy} & ROC-AUC & Accuracy & ROC-AUC & Accuracy \\
         \hline
         \multicolumn{2}{c|}{Min-weight masking} & 0.9804 & 0.9597 & 0.9010 & 0.8807 \\
         \multicolumn{2}{c|}{Max-weight masking} & 0.9663 & 0.9516 & 0.8843 & 0.8781 \\
         \multicolumn{2}{c|}{Random masking} & 0.9464 & 0.9032 & 0.9007 & 0.8786 \\
         \hline
    \end{tabular}}
\end{table}

\subsection*{Hyperparameter optimization}\label{subsec2}
As multiple fully-connected layers were used for transfer learning to downstream tasks, we optimized the number of fully-connected layers, the number of units in each layer and the learning rate. For efficiency, we performed parameter optimization on the 619 pMHC/TCR pairs dataset so that our model achieved better performance in predicting TCR and antigen-peptide binding specificity. In addition to the softmax output layer, we tested 2, 3, and 4 hidden layers. The number of units in each layer is enumerated in Table \ref{table:2}. We found that 3 fully connected layers with 128, 63, and 16 units in downstream tasks achieved the best performance. The learning rate is tuned to 0.001.

\begin{table}[]
\centering
\caption{Hyperparameter tuning of the fully-connected layers for downstream task}
\label{table:2}
\begin{tabular}{cc|c|c}
\hline
\multicolumn{2}{c|}{Network architecture}  & ROC-AUC & Accuracy\\ \hline
\multicolumn{1}{c|}{\multirow{3}{*}{2 hidden layer}} & 64,8  & 0.9619 & 0.9435 \\
\multicolumn{1}{c|}{}  & (64,16)  & 0.9032  & 0.9464  \\
\multicolumn{1}{c|}{}  & (64,32) & 0.966 & 0.9596 \\ \hline
\multicolumn{1}{c|}{\multirow{3}{*}{3 hidden layer}} & (128,64,8) & 0.977 & 0.9434 \\
\multicolumn{1}{c|}{} & (128,64,16) & \textbf{0.9804} & \textbf{0.9597} \\
\multicolumn{1}{c|}{}  & (128,64,32) & 0.9757  & 0.9354\\ \hline
\multicolumn{1}{c|}{\multirow{4}{*}{4 hidden layer}} & 128,64,16,8  & 0.9713 & 0.9511 \\
\multicolumn{1}{c|}{}  & (128,64,32,4)  & 0.9694  & 0.9513 \\
\multicolumn{1}{c|}{}  & (128,64,32,8)  & 0.977 & 0.9516 \\
\multicolumn{1}{c|}{}  & (128,64,32,16) & 0.9681  & 0.9354 \\ \hline
\end{tabular}
\end{table}
\subsection*{Performance comparison}\label{subsec2}
Next, we compared our method with three mainstream prediction models for TCR-epitope binding, including pMTnet~\citep{rf17}, TCRex~\citep{rf34} and NetTCR~\citep{rf16}. We conducted comparison experiments on the small dataset with 619 TCR-pMHC bindings. pMTnet can predict all 619 TCR-epitope pairs. However, TCRex and NetTCR supported only  a portion of MHC alleles. For example, NetTCR only supported the HLA-A: 0201 allele and the antigenic peptides shorter than 10 amino acids. So, we can only selected a portion of samples as the test set. For TCRex, 163 out of 619 TCR-pMHC pairs were applicable. For NetTCR, only 59 out of 619 TCR-pMHC pairs were applicable. As a result, apart from the 619 dataset, two subsets were established for performance evaluations. Table \ref{table:3} shows the performance of different methods on these datasets.

\begin{table*}[!htpb]
\centering
\caption{Performance comparison of ATMTCR to three competitive methods on \textit{small} dataset. '-' indicated the method did not support the corresponding dataset.}
\label{table:3}
\begin{tabular}{c|cccc}
\hline
 Method & All 619 pairs & 163 out of 619 pairs & 59 out of 619 pairs \\ \hline
pMTnet  & 0.827 & \textbf{0.894}& 0.932 \\
NetTCR & - & -   & 0.528\\
TCRex & - & 0.625 & -  \\ \hline
ATMTCR (min-weight masking) & \textbf{0.9804} & 0.7820  & \textbf{0.9405} \\
ATMTCR (max-weight masking) & 0.9663 &0.7782  & 0.9059\\
ATMTCR  (random masking)     & 0.9464&0.7105  & 0.9135  \\ \hline
\end{tabular}
\end{table*}

On all 619 pairs, our method performed entirely better than pMTnet, across each masking strategy. On the 163 subsets, our method was compared to pMTnet and TCRex. We found our method performed slightly inferior to pMTnet, but still much better than TCRex. On the 59 subset, our method was compared to pMTnet and NetTCR. The ROC-AUC values achieved by each masking strategy was much higher than that of NetTCR. Compared to pMTnet, the performance obtained by min-weight masking strategy was better than that of pMTnet, and other masking strategy-derived performance were also comparable to pMTnet.

\subsection*{Attention revealed important residues affecting TCR-pMHC binding}\label{subsec2}
\begin{figure}[htbp]  
	\captionsetup{labelformat=simple, position=top}
	\centering
	\subfloat[\textit{large} dataset]
	{
		\begin{minipage}[b]{.5\columnwidth}
			\centering
			\includegraphics[width=4cm]{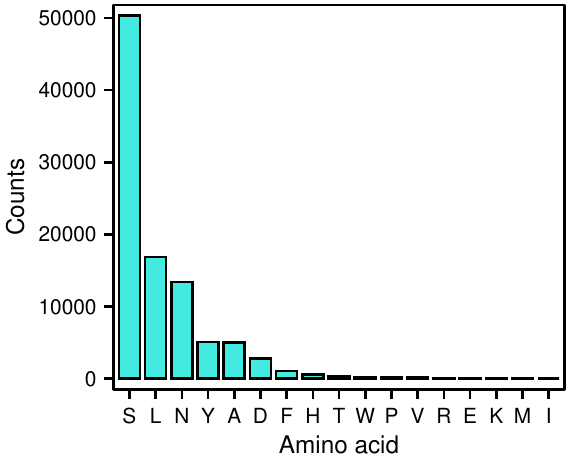}
	\end{minipage}}
	\subfloat[\textit{small} dataset]
	{
		\begin{minipage}[b]{.5\columnwidth}
			\centering
			\includegraphics[width=4cm]{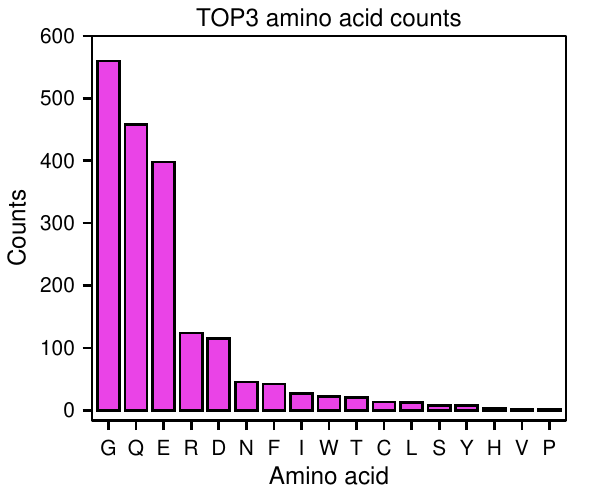}
	\end{minipage}}
	\caption{Frequency distribution of top 3 amino acids with highest attention weights among TCR CDR3 sequences on two datasets. }	\label{fig:top3stat}
\end{figure}

To explore which amino acids play a decisive role in TCR-antigen binding, we applied the learned attention weights to screen for important amino acids. After pre-training and transfer learning on the two downstream datasets, the top 3 amino acids with the highest attention weights were selected regarding each CDR3 sequence, and then the cumulative frequency was calculated. Figure \ref{fig:top3stat} showed the frequency histograms of top-3 amino acids on two datasets. It can be found that although the frequency distributions differed from each other, most amino acids overlapped between these two datasets. For example, the occurrence frequency of asparagine (N), phenylalanine(F), aspartic acid (D), threonine(T) ranks within top 10 in both datasets.

We found quite a few previous studies that reported the importance of these four amino acids in the function of TCR binding to antigens. For example, Lin X et al.~\citep{rf35} have reported that the crystal structure of TCR 2B4 contains many threonine (T) and asparagine (N) residues in the CDR loop region. The 3-dimensional crystal structure of TCR 2B4 showed that, when binding to antigen peptides, the alanine (A), threonine (T), and asparagine (N) residues have high Corey-Pauling-Koltun (CPK) representation~\citep{rf36}. This consistence of our result to previous study at least partially validate the effectiveness of our model. Waldt et al. ~\citep{Waldt2018} demonstrate that the serine/threonine kinase Ndr2 is critically involved in the initiation of TCR-mediated LFA-1 activation with open conformation in T cells. This consistence of our result to previous study at least partially validate the effectiveness of our model.\par

\begin{figure*}[htbp]  
	\captionsetup{labelformat=simple, position=top}
	\centering
	\subfloat[Glycine(G)]
	{
		\begin{minipage}[b]{.6\columnwidth}
			\centering
			\includegraphics[width=4.5cm]{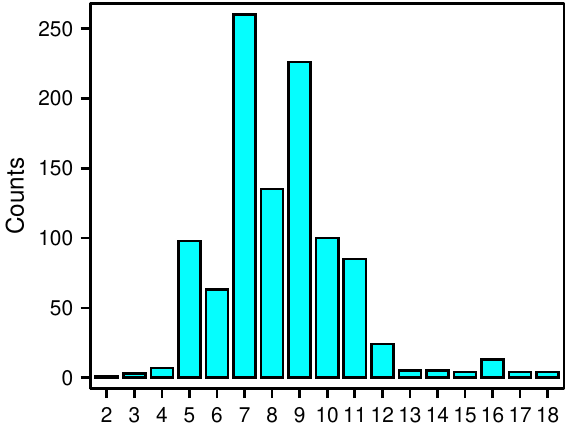}
	\end{minipage}}
	\subfloat[Glutamine(Q))]
	{
		\begin{minipage}[b]{.6\columnwidth}
			\centering
			\includegraphics[width=4.5cm]{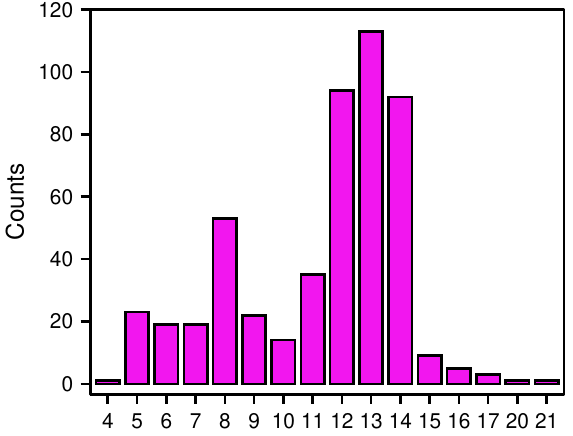}
	\end{minipage}}
	\subfloat[Glutamic acid(E)]
	{
		\begin{minipage}[b]{.6\columnwidth}
			\centering
			\includegraphics[width=4.5cm]{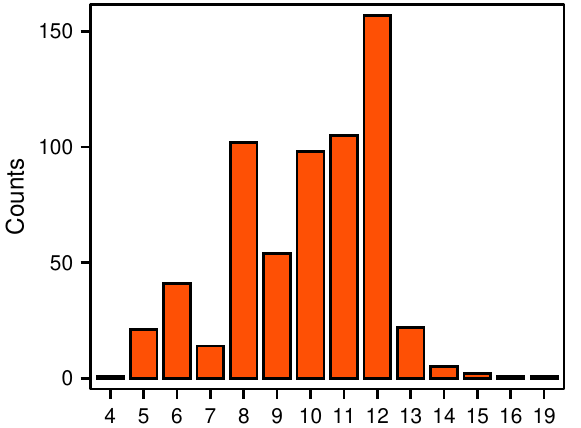}
	\end{minipage}} \\
	\subfloat[Serine(S)]
	{
		\begin{minipage}[b]{.6\columnwidth}
			\centering
			\includegraphics[width=4.5cm]{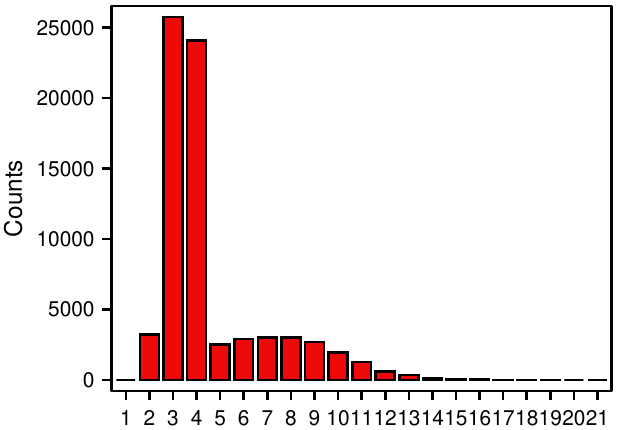}
	\end{minipage}}
	\subfloat[Leucine(L)]
	{
		\begin{minipage}[b]{.5\columnwidth}
			\centering
			\includegraphics[width=4.5cm]{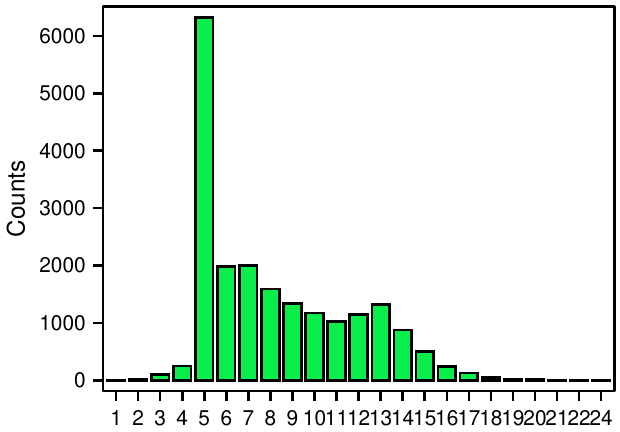}
	\end{minipage}}
	\subfloat[Asparagine(N)]
	{
		\begin{minipage}[b]{.6\columnwidth}
			\centering
			\includegraphics[width=4.5cm]{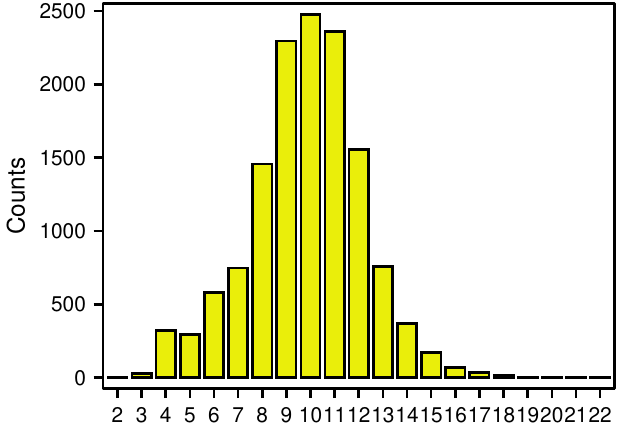}
	\end{minipage}}
	\caption{Distribution of positional preference of three amino acids with highest cumulative attention weights on two datasets.  On the \textit{small} dataset (upper row) and \textit{large} dataset (bottom row), three amino acids with highest cumulative attention weights were glycine(G) and glutamine(Q) and glutamic acid (E), serine (S) and leucine (L) and asparagine (N), respectively.}	\label{fig:position}
\end{figure*}

We also explored the positional preference of top-3 amino acids in two datasets, as shown in Figure \ref{fig:position}. Lu et al.~\citep{rf17} divided the TCR CDR3 sequence into 6 equal-length fragments, plotted the trend of the binding affinity of the residues to antigen peptides, and concluded that the residues in the middle region of CDR3 sequences were more likely to affect the binding specificity.  Our statistical results was consistent with the conclusions of previous study.

\section*{Discussion }\label{sec3}
Understanding the binding of TCR and antigenic peptides plays a key role in the estimation of the efficacy of immune checkpoint inhibitors and development of immunotherapies. However, due to the high polymorphism of TCR sequences, it is difficult to capture the biological mechanism of antigenic peptides eliciting T cells. On the other hand, the incredible diversity of TCR repertoire make it impractical to screen TCR-antigen bindings via wet-lab experiments.

Model-based prediction of TCR-antigen binding also run into difficulty due to the paucity of labeled data. Self-supervised learning showed promising progress in learning informative representations by designing some pretext tasks. We adopted this new technique to perform representation learning of TCR sequences, and meanwhile tied to increase the interpretability of the model by using self-attention mechanism.

As a result, we observed interesting appearance and distribution patterns of some amino acids that potentially determine TCR-epitope binding specificity. We have drawn the conclusion that specific amino acids localized in key region of the CDR loops would bind strongly to antigenic peptides and MHC molecules, thereby eliciting immune response. Lin X et al.~\citep{rf35} have delineated the 3Dl crystal  structure of TCR-pMHC complex, and showed some regions were more attractive to antigenic peptide. From this perspective, we hope to extend the model in our future work to narrow the region where TCR binds to antigenic peptides for triggering immune responses, thereby reducing the experimental cost.\par
In this study, we performed self-supervised representation learning on a large number of TCR sequences, and fine-tuned the model on experimentally validated TCR-epitope data. Our experimental verified our model achieved promising results. By pretraining on a variety of TCR sequences, the model allows us to better understand the determinant factors of TCRs in binding to epitopes, as well as the importance of the HLA type for the presentation of epitopes. Given more data, we believe our model is possible to predict whether TCRs recognize previously unseen epitopes.

\section*{Reference}
\bibliographystyle{unsrt}
\bibliography{ref}

%
%
%


\end{document}